\documentclass[%
 aip,
 amsmath,amssymb,
 reprint,%
]{revtex4-1}

\usepackage{color}
\usepackage{graphicx}
\usepackage{epsfig}
\usepackage{dcolumn}
\usepackage{bm}

\usepackage[utf8]{inputenc}
\usepackage[T1]{fontenc}
\usepackage{mathptmx}
\usepackage{etoolbox}

\usepackage[normalem]{ulem}

\begin{document}
\title{Denjoy's anachronistic topological viewpoint on Aubry transition}

\author{O. C\'epas}
\affiliation{Institut N\'eel, CNRS, Universit\'e Grenoble Alpes, Grenoble, France}
\author{G. Masbaum}%
\affiliation{ Sorbonne Université, Université Paris Cité, CNRS, Institut de Math\'ematiques de Jussieu - Paris Rive Gauche (IMJ-PRG), F-75005 Paris, France }
\author{P. Qu\'emerais}
\affiliation{Institut N\'eel, CNRS, Universit\'e Grenoble Alpes, Grenoble, France}

\date{\today}

\begin{abstract}
  The Aubry transition is a phase transition between two types of
  incommensurate states, originally described as a transition by
  ``breaking of analyticity''.  Here we present Denjoy's
  (anachronistic) viewpoint, who almost a hundred years ago described certain mathematical
  properties of circle homeomorphisms with irrational rotation numbers.
  The connection between the two lies in the existence of a change of
  variables from the incommensurate ground state variables to new
  simple phase variables that rotate by a constant irrational angle.
  This confers a cyclic order, an essential property of models with
  the Aubry transition.  Denjoy's description indicates that there are
  two types of cyclic order, distinguished by the regular or singular
  nature of the change of variables or, in mathematical terms, by the
  distinction between topological conjugacy versus semiconjugacy. This allows
  rephrasing the breaking of analyticity as a breaking of topological
  conjugacy.  We illustrate this description with numerical
  calculations on the Frenkel-Kontorova model.
\end{abstract}
 
  \maketitle
 
 { \quotation{Incommensurate structures are ubiquitous and very important in condensed matter physics. It was in 1978 that S. Aubry demonstrated, within the framework of the Frenkel-Kontorova model, the existence of two types of incommensurate phases with very different physical properties. Although known within the physics community, Aubry's theory is often misunderstood. In the original papers, it is mainly based on the mathematical study of the standard map trajectories whose complexity is significant. In the present paper, we describe a much simpler, though little known approach. It was introduced by V. Bangert in 1988 and is based on the existence of a circle homeomorphism that relates the positions of successive atoms in the incommensurate structure. This allows the use of earlier results obtained by H. Poincar\'e and A. Denjoy at the beginning of the previous century. This approach allows a great simplification and leads directly to Aubry's results: the existence of two types of incommensurate phases as well as the emergence of a Cantor set to which the atomic positions belong in one of the two phases.}}

\section{Introduction}

The Aubry transition, first published in 1978~\cite{Aubry}, remains a
surprising discovery in the field of phase transitions. It does not
describe a conventional phase transition that
spontaneously breaks a symmetry of the model. Instead, it describes a
transition between two incommensurate phases, a "sliding" phase, which possesses a continuous
symmetry that the model explicitly breaks away from an integrable
point, and a standard "pinned" phase that has no more or less symmetry
than the model itself. The presence of such an additional continuous
symmetry is somewhat remarkable. The emergence of physical properties
in general, and the emergence of symmetries in particular, notably in
the thermodynamic limit, sometimes called accidental
symmetries~\cite{acc}, are now central to condensed matter physics. In
the presence of frustrating interactions, degenerate manifolds of
ground states may exist giving some apparent symmetry. For example, the
magnetic 6-state clock model which explicitly breaks the spin rotation
symmetry has a phase at intermediate temperatures which has the
rotation symmetry~\cite{Jose}.  Despite these rather vague analogies,
the Aubry transition seems rather singular. Nearly 50 years after its
discovery, it is still regarded as a separate chapter of the physics
of phase transitions.

The transition has been theoretically described in various one-dimensional models, ranging from classical models to quantum models of charge density waves. The essential common property of these models, which in itself defines a class of models, appears to be the presence of ``cyclic order'' in the ground states. This property, proven from Aubry-Le Daeron's \textit{fundamental lemma}~\cite{aubry_ledaeron} for the Frenkel-Kontorova model, is discussed in the mathematical literature~\cite{MacKay,Meiss,Gole}, but, although numerically observed, remains unproven for other physical models~\cite{Ledaeron-aubry-holstein,CQ}. The Aubry transition does not have so far generalizations in higher dimensions or to
other symmetries. Seeking alternate viewpoints may help to understand its real extent.
Nevertheless, its occurence, originally in the simple Frenkel-Kontorova model, has permitted some major breakthroughs in nonlinear dynamics, particularly the Aubry-Mather sets~\cite{Aubry,mather}  of the standard map, also called cantori~\cite{Percival_cantori} (see for example Ref.~\onlinecite{Moser}). When described in dynamical terms, the Aubry transition is sometimes called the stochastic transition~\cite{Dzyalo}, as it is concomitant with the onset of chaos, which has sometimes led to the mistaken belief that the ground states of the Frenkel-Kontorova model could be chaotic~\cite{Bak}.

There are two striking features in the Aubry transition: first the persistance of the incommensurate sliding phase away from integrability which we mentioned above. This is related to the KAM theorem as usually applied to the standard map. Second, the existence of self-similar patterns in the pinned phase. This is the consequence of the existence of Aubry-Mather Cantor sets or cantori. To understand physically why these two properties occur, it is useful to adopt the anachronistic perspective of Denjoy's 1932 results~\cite{Denjoy} as used by Bangert~\cite{bangert} in its analogy with geodesics on the torus and emphasize the topological nature of the transition.

The ground states of this class of models appear to have a form of topological rigidity when the parameters of the model are varied: the continuous variables $x_n$ of the ground states are constrained by equations of the form,
\begin{equation}
x_{n+1}=F(x_n), \label{R0}
\end{equation}
where $F$ are certain functions from $\mathbb{R}$ to $\mathbb{R}$ (see section~\ref{section3}); they are lifts of circle homeomorphisms $f$ (and hence have special properties: they are continuous, invertible and the inverse is also continuous).  A physical consequence is cyclic order, a property that is conserved throughout the phase diagram provided that the model satisfies certain specific properties. This cyclic order imposes constraints on the ground states : for example the configurations must follow some special sequences (Sturmian words~\cite{lothaire}) which is a form of hidden order.
It is a direct consequence of the topological classification of circle homeomorphisms by Poincaré at the end of the 19$^{\mbox{th}}$ century. A circle homeomorphism is, indeed, ``very similar'' to a rotation and admits a rotation number $\rho$ which is either a rational or an irrational number. This rotation number illustrates the fact that  $f$ behaves on average like a rotation by an angle $\rho$ and $F$ as a translation of step $\rho$. In the Frenkel-Kontorova model, a rational rotation number corresponds to a commensurate structure, while an irrational number corresponds to an incommensurate structure and it is only in the incommensurate case that the Aubry transition takes place.

When $F$ (and $f$) has an irrational rotation number, Poincar\'e showed that
there exists a change of variables $H$ to new phase variables
\begin{equation} \phi_n=H(x_n), \end{equation}
where $\phi_n$ increase by a constant irrational number $\rho$:
\begin{equation} \phi_n=n \rho+ \phi_0. \end{equation}
The ground state, considered in terms of the new variables $\phi_n$ is a one-dimensional regular lattice with lattice constant $\rho$, whereas viewed in the original variables $x_n$, it is deformed by $H$ in a manner that lies at the heart of the Aubry transition.
The change of variables is expressed by the equation
\begin{equation} H \circ F= R_{\rho} \circ H, \label{semiconj} \end{equation}
where $H$ is in general a \textit{semiconjugacy} (a monotonically increasing, but not necessarily strictly increasing, continuous function) and $R_{\rho}$ a rotation of irrational angle $\rho$, $R_{\rho}(x)=x+ \rho$. 
$F$ and $R_{\rho}$ are in some sense ``equivalent'' up to a change of variables $H$. 

Denjoy studied in 1932~\cite{Denjoy} the regularity properties of $H$ and what determines them for a given $F$. He showed what is now known as Denjoy's theorem~\cite{Milnor,Katok,Yoccoz}, that if $F$ is sufficiently regular (of class C$^2$ - slightly less restrictive conditions also exist), then $H$ is necessarily strictly increasing, hence a homeomorphism, hence it is invertible. In this case, $H$ is a topological \textit{conjugacy}~\cite{note}, and Eq.~(\ref{semiconj}) becomes \begin{equation} F=H^{-1} \circ R_{\rho} \circ H. \label{conj} \end{equation} Furthermore, Denjoy also constructed some circle homeomorphisms $F$ of class $C^1$, known as Denjoy's counterexamples~\cite{Milnor,Katok,Yoccoz} which are not conjugate to a rotation, but remain semiconjugate. In this case, $H$ is still monotonically increasing but not strictly so that $H$ is not invertible and only Eq.~(\ref{semiconj}) holds.

The difference between (\ref{semiconj}) with $H$ non invertible and (\ref{conj}) is of crucial importance and has physical consequences. 
The Aubry transition is precisely a change in the properties of the function $H$: $H$ is a conjugacy function (a regular function) in the sliding phase and a semiconjugacy function (a singular Cantor function) in the pinned phase.
The difference is the following.
While, on one hand, the angular variables $n \rho + \phi_0$~mod~1, $\rho$ irrational, are dense in $[0,1]$ (on a circle), a topological conjugacy ensures that the original variables $x_n$~mod~1 are also dense on $[0,1]$: the topology of the circle is not changed by the homeomorphism $H$. On the other hand, a semiconjugacy generally has gaps on the circle (i.e. the variables $x_n$~mod~1 are no longer dense in $[0,1]$).
This allows one to rephrase the Aubry transition as a breaking of topological conjugacy, i.e. a change of topology of the space of the ground state circular variables $x_n$~mod~1, from a whole circle (sliding phase) to a circle with gaps (pinned phase) which is a cantorus.

\section{Models and hypotheses}

We consider the classical energy to minimize,
\begin{equation}
\label{newenergy}
E(\{x_n\}) = \sum_n L(x_{n+1},x_n), 
\end{equation}
where $x_n$ are some real variables, e.g. atomic positions~\cite{ft1}. The system is infinite and is assumed to have a discrete translation symmetry with period one: $x_n \rightarrow x_n+1$ does not change the energy. 
Typically, the energy could be the Frenkel-Kontorova model,
\begin{equation}
\label{energy}
E(\{x_n\}) = \frac{1}{2} \sum_n (x_{n+1}-x_n-a)^2 + \frac{K}{(2\pi)^2} \sum_n \cos(2 \pi x_n),
\end{equation}
where $a$ and $K$ are some parameters. The first term is the elastic energy of the first nearest neighbors which is minimized when the bond lengths equal to $a$: we consider no terms such as longer range elastic couplings $(x_{n+2}-x_{n})^2$ which may be frustrated and change the physics~\cite{Axel}. The second term is a potential with periodicity one.
The model has thus a continuous translation symmetry at $K=0$ which becomes a discrete translation symmetry at $K>0$.

The function $L$ is assumed to satisfy several other properties (see note~[\onlinecite{notePropertiesL}]) which are essential to ensure the validity of Aubry-Le Daeron's fundamental lemma~\cite{aubry_ledaeron}, which itself ensures the existence of $F$. 

The Aubry transition is often discussed on the basis of a two-dimensional dynamical system, which is a symplectic twist map~\cite{Gole}.
Indeed, the extremalization of $E$ for the Frenkel-Kontorova model gives the equation for the equilibrium of forces 
\begin{equation}
 x_{n+1}+x_{n-1} -2x_n + \frac{K}{2\pi} \sin(2 \pi x_n)=0.
  \label{firstderivative}
\end{equation}
By introducing the bond lengths $\ell_n \equiv x_{n+1}-x_n$, 
\begin{equation}
  \left\{
  \begin{array}{ll}
  x_{n+1} &= x_{n}+\ell_{n},   \\
  \ell_{n+1} &= \ell_{n}- \frac{K}{(2\pi)} \sin(2 \pi x_{n+1}). \end{array} \right.
  \label{ds}
\end{equation}
This defines the two-dimensional standard map where $n$ is seen as a discrete time~\cite{Aubry,aubry_ledaeron,mather}. 
 A modification of the potential gives further harmonics to the sin term and modifies some characteristics of the Aubry transition~\cite{CQ2} but preserves the two-dimensional character, which arises from the special choice $L(x_{n},x_{n+1})$. Note however that other models~\cite{Ledaeron-aubry-holstein,CQ} with Aubry transition do not fit the form of Eq.~(\ref{newenergy}) so that the two-dimensional character of the dynamics, as in the Frenkel-Kontorova model, is not a prerequisite. 
 The physics of the standard map is remarkable and now well documented, and its trajectories, dynamically stable or unstable, have been extensively studied. The key point in Aubry's theory was to consider, among all trajectories of the standard map, those that are physically stable (which are not necessarily those which are dynamically stable). This led to the definition of what constitutes a minimal energy trajectory, thus resolving the confusion between dynamic stability and physical stability. The ground states can be found in chaotic regimes without being physically unstable themselves.
 
 \section{Underlying circle map constraining the ground states}
 \label{section3}

\subsection{Non-crossing minimal energy states} 

In the approach developped by Aubry and Le Daeron~\cite{aubry_ledaeron}, one first defines the minimal energy states for infinite chains and derives several exact properties given the hypothesis of the model.
These properties which we recall are defined in a series of publications~\cite{aubry_ledaeron,bangert,Gole}.
The minimal energy states are not necessarily the absolute ground states, they may have higher energies. They are \textit{minimal} energy states in the sense that any deformation with fixed boundaries of any segment of the chain will increase the energy.

Several general results have been obtained:

\begin{itemize}
\item When the minimal energy state $x_n$ is viewed as a physical trajectory over the discrete time $n$, it happens that two such trajectories cannot cross more than once. This is the fundamental lemma of Aubry-Le Daeron~\cite{aubry_ledaeron}. This property is the same for geodesics of class A as noticed by Chenciner~\cite{Chenciner}. Minimizing the action [Eq.~(\ref{newenergy})] gives broken geodesics (in the plane $x_n,n$)  and one needs some special properties~\cite{notePropertiesL} of $L$ to avoid these geodesics to cross more than once~\cite{bangert}. These properties of $L$ are supposed to be satisfied here.
\item By using the discrete translation symmetry of the model, one can construct an ensemble of degenerate states $\mathcal{E}$ as follows. For a given minimal energy state, $x_n$, the translated state $x_n+p$ where $p$ is an integer is also a minimal energy state. Furthermore, it is always possible to consider a new state with a renumbering of the sites and translation, $x_n'=x_{n+q}+p$ for any integers $p,q$. It means that it is always possible to ``start'' at any point $q$ of a given state and consider it as the ``first'' point 0 of a new state $x_0'$ and take it in $[0,1]$ by an appropriate choice of translation $p$.
Furthermore, we can even take limits of such states.
The resulting ensemble of states $\mathcal{E}$ is closed in the appropriate (weak) topology\cite{aubry_ledaeron}. Aubry-Le Daeron's fundamental lemma implies that $\mathcal{E}$ is totally ordered~\cite{bangert,Gole}. It means that, for two states $\{ x_n \}$ and $\{ x_n' \}$ in $\mathcal{E}$, if for a given $n$, $x_n<x_n'$ then $x_{n+k}<x_{n+k}'$ for all integers $k$.
If $n$ is viewed as a discrete time, these two trajectories never cross. \textit{Any two states in this ensemble of degenerate minimal energy states never cross.}
In particular,
\begin{equation}
  \text{if} \  x_n<x_n'  \ \mbox{then} \ x_{n+1}<x_{n+1}' \label{Greg}
\end{equation}
  which is the key for what follows.

\end{itemize}

\subsection{Effective rotation} \label{effectiverotation}

Bangert~\cite{bangert} used Aubry-Le Daeron's fundamental lemma~\cite{aubry_ledaeron} and the ordering of the ensemble of degenerate states $\mathcal{E}$ considered above to prove that a minimal energy state $\{x_n \}$ obeys the equation
\begin{equation}
x_{n+1}=F(x_n), \label{R}
\end{equation}
where $F$ is a function with some important properties explained below (examples of $F$ are given in Figs.~\ref{fig1a} and \ref{fig1b}). This equation is central to the present discussion and important to understand physically because it constraints the minimal energy states. The function $F$ is first defined  on the special points $x_n$  that belong to  some state $\{ x_n\}$ in $\mathcal{E}$ (the same $F$ works for all the states in $\mathcal{E}$). 
It can be extended to a function from  $\mathbb{R}$ to  $\mathbb{R}$ possibly with some arbitrariness which does not change the properties expressed below.

One may wonder why it is possible to  define such a function $F$ which after all means to reduce an iteration equation which involves $x_{n+1}$ as a function of $x_n$ and $x_{n-1}$ [Eq.~(\ref{firstderivative})] to an iteration involving only $x_n$ [Eq.~(\ref{R})]?
If one only looks at Eq.~(\ref{firstderivative}), it does not seem clear that if two points of an orbit were the same, say $x_n=x_p$ (modulo 1), $n \neq p$, it would give the same following point $x_{n+1}=x_{p+1}$ (the condition for $F$ to be a well-defined function). However, the fundamental lemma and the noncrossing of all translated and renumbered orbits implies that $x_n=x_p$, $n \neq p$ is impossible unless the orbit and its translated (or renumbered) orbit are the same, thus giving indeed the same following point $x_{n+1}=x_{p+1}$ and allowing to define a function $F$.

The function $F$ has several properties (all the details and rigorous arguments are given in Ref.~\onlinecite{bangert}):
\begin{itemize}
\item  $F$ is such that $F(x+1)=F(x)+1$ because of translation symmetry (translating a configuration by 1 leads to a new configuration with the same energy which must satisfy the same iteration process).
\item $F$ is a strictly increasing bijection, as follows from (\ref{Greg})~\cite{bijection0}.
\item As already mentioned, $F$ can be continuously extended to a function from $\mathbb{R}$ to $\mathbb{R}$. Indeed either the $x_n$~mod~1 are dense in $[0,1]$ and the extension is obvious or they are not and a linear interpolation is sufficient to define $F$ everywhere in $[0,1]$ and hence in $\mathbb{R}$.
\end{itemize}
$F$ is then a homeomorphism from $\mathbb{R}$ to $\mathbb{R}$.  It is the lift of a circle homeomorphism. Recall that the circle $S^1$ can be thought of as the interval $[0,1]$ with the end points identified.
  Given a function $F$ as above, the function from $\mathbb{R}$ to $[0,1[$,
      \begin{equation} f \equiv F~\mbox{mod}~1, \end{equation} satisfies $f(x+1)=f(x)$, so is periodic with period 1. By considering angular variables $x \in [0,1[$, one can consider $f$ as a map from $S^1$ to $ S^1$. This  $f$ is a circle homeomorphism and $F$ is its lift to $\mathbb{R}$ satisfying $F(x+1)=F(x)+1$. We will sometimes blur the distinction between $F$ and $f$. For example, we will use the notation $R_{\rho}$ both for the translation $F(x)=x+\rho$ and the corresponding circle rotation of angle $\rho$.

            Thus the minimal energy states of the general model  given by Eq.~(\ref{newenergy}) satisfy the strong constraint, Eq.~(\ref{R}), where $F$ is a lift of a circle homeomorphism. We now recall some consequences for the minimal energy states.

\subsection{Constraints for the minimal energy states}
\label{constraints}

We consider here in general the dynamical process which consists of iterating
\begin{equation} x_{n+1}=F(x_n), \label{Rbis}
\end{equation}
starting from an initial condition $x_0$, 
where the function $F$ here is any homeomorphism from $\mathbb{R}$ to $\mathbb{R}$ with $F(x+1)=F(x)+1$. The orbits satisfy several properties, as originally shown by Poincaré and recalled in different places~\cite{bangert,MacKay,Gole,Katok,Arnold}.

A first property is that, under iterations,
\begin{equation}
  \mbox{lim}_{n \rightarrow + \infty} \frac{x_n-x_0}{n} = \rho,
\end{equation}
where $\rho$ is a real number, the rotation number. It means that 
\begin{equation} x_n \sim x_0+ n \rho \end{equation} for large $n$. The rotation number $\rho$ is independent of $x_0$ and is a characteristic of $F$.  It can be a rational or irrational number.  Poincaré proved that when $\rho=r/s$ is rational, there exists a point $x$ for which $F^{(s)}(x)=x+r$, where $F^{(s)}$ is the function $F$ iterated $s$ times.
 In other words there are some orbits which satisfy $x_{n+s}=x_n+r$, they are periodic orbits when taken modulo 1. For other $x$, the orbit needs not be periodic but tends asymptotically to a periodic orbit. When $\rho$ is irrational, the orbit is always nonperiodic.

In fact, the variables $x_n$ are very much constrained since one can further show that
\begin{equation}
|x_n-x_m-(n-m) \rho| < 1, \label{inequality}
\end{equation}
for all $n$ and $m$ (see \textit{e.g.} Ref.~\onlinecite{Gole}). In particular, we have $|x_{n}-x_0-n \rho| < 1$.

If $F$ comes from the Frenkel-Kontorova model, as in \ref{effectiverotation}, this shows that each minimal energy state of that model can be characterized by a rotation number $\rho$ (the averaged lattice spacing) and cannot be far from the undistorted lattice with lattice spacing $\rho$. In fact, for the class of Frenkel-Kontorova models, the inequality (\ref{inequality}) was directly proved already by Aubry-Le Daeron~\cite{aubry_ledaeron} without using circle homeomorphisms.

\begin{figure}[h]
\psfig{file=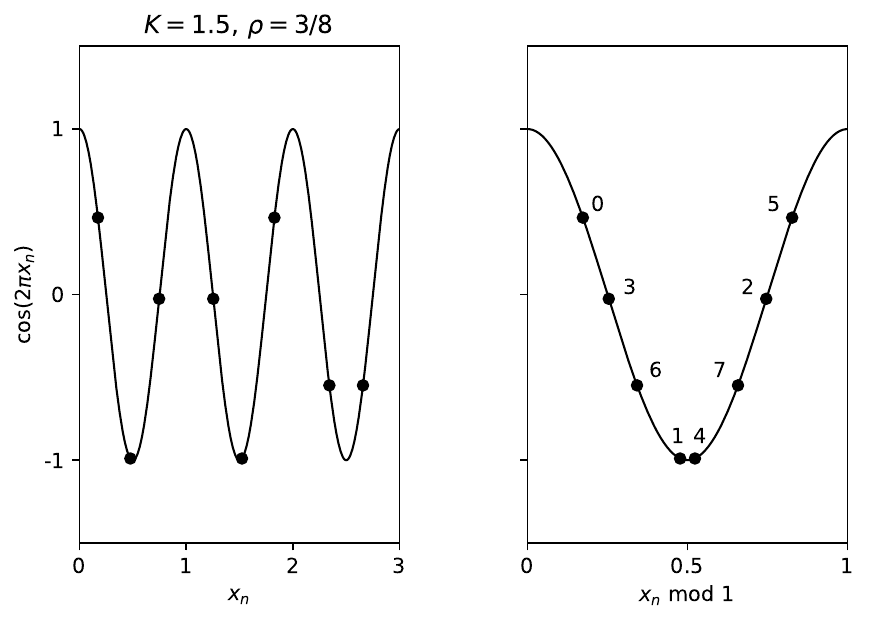,width=8.8cm,angle=-0}
\psfig{file=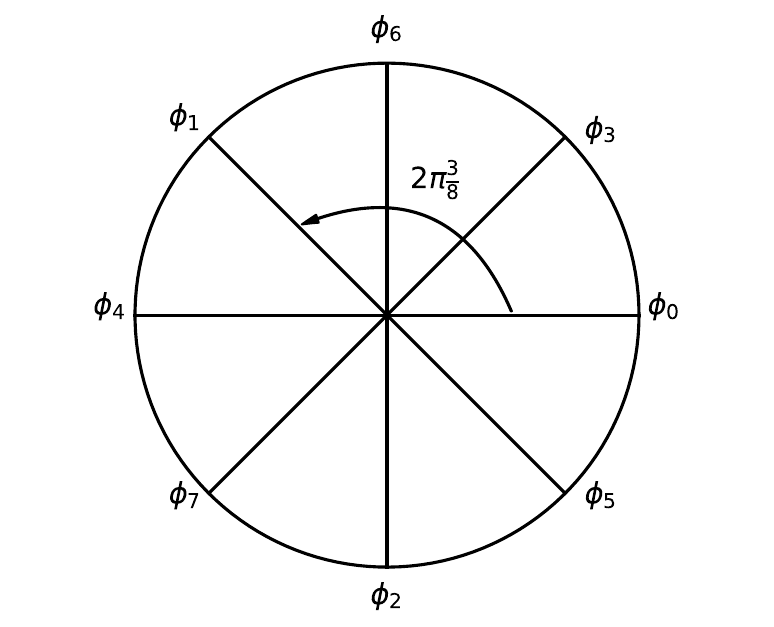,width=7cm,angle=-0}
  \caption{Periodic ground state configuration $\{x_n\}$ of the Frenkel-Kontorova model for $\rho=3/8$ and $K=1.5$ (top left). The order of the $x_n~$mod~1 variables (top right) is the same as that of the cyclic order of the rigid rotation with the same $\rho$ (bottom).}
\label{figcyclic}
\end{figure}

A second property is cyclic order which is the special order  of rigidly rotated points on the circle (see Fig.~\ref{figcyclic} bottom for the example of a rigid rotation with $\rho=3/8$).
An orbit of $F$ which is periodic or has an irrational rotation number also has cyclic order.
It means that the angular variables $x_n~$mod~1 are arranged in the same way as that of the angular variables of the rigid rotation with the same $\rho$, $\phi_n=n \rho~$mod~1~\cite{bijection}.
In fact, the cyclic order for an irrational rotation number is described by an infinite Sturmian word, a sequence of 0 and 1 given for example by $[(n+1)\rho+\alpha]-[n\rho+\alpha]$ where $[...]$ is the integer part and $\alpha$ is a phase~\cite{lothaire}. As a consequence a ground state is also described by the same Sturmian word (with an appropriate choice of phase), called a ``uniform'' state in the physics literature~\cite{Ducastelle}.

We illustrate this property for the ground states of the Frenkel-Kontorova model. For example consider a rational $\rho=r/s=3/8$.  The commensurate ground states $x_n$ with that $\rho$, i.e. satisfying $x_{n+s}=x_n+r$, are obtained numerically by minimizing the energy by a standard gradient descent algorithm. One of them is represented in Fig.~\ref{figcyclic}, top left: it is sufficient to represent the first unit-cell of size $r$ which contains $s$ atoms with positions $x_n$ ($n=1, \dots, s$). The ordering of the $x_n~$mod~1 variables (top right) is precisely that of the sequence $\phi_n$ (bottom). For example $x_1~$mod~1 must be in between $x_6~$mod~1 and $x_4~$mod~1 just as $\phi_1$ is in between $\phi_6$ and $\phi_4$. This is a strong constraint and a hidden form of order in the ground states of the Frenkel-Kontorova model [or any model of the form given by Eq.~(\ref{newenergy})], and a direct consequence of Eq.~(\ref{R}).

We are now interested in $\rho$ irrational for which the Aubry transition takes place.

\subsection{Numerical construction of $F$}
\label{num}

We construct numerically examples of $F$ functions for the Frenkel-Kontorova model when $\rho$ is irrational. When $\rho$ is irrational, the Frenkel-Kontorova model has an Aubry transition at a critical value of $K=K_c$ which depends on $\rho$~\cite{Aubry}. For $K<K_c$, the phase is sliding and the ground state angular variables $x_n$~mod~1 take all values in $[0,1]$, whereas, for $K>K_c$, the phase is pinned and not all values are allowed. This is most apparent in Aubry's ``hull'' function which is continuous for $K<K_c$ and discontinuous for $K>K_c$, hence the breaking of analyticity~\cite{Aubry}. Here we show the incipient traces of the Aubry transition in $F$.

\begin{figure}[h]
    \psfig{file=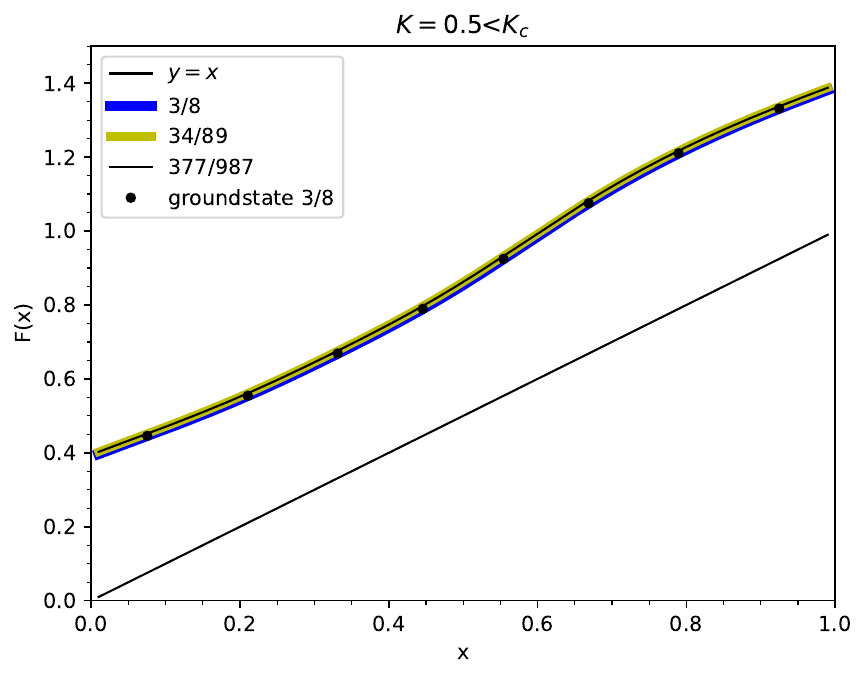,width=8.7cm,angle=-0}
  \caption{The function $F$ constraining the ground states [see Eq.~(\ref{R})] for the Frenkel-Kontorova model for $K<K_c$ (sliding phase). It converges to a single function $F$ when the sequence of rational approximants (only three are given here) converges to the irrational number $\rho=(3-\sqrt{5})/2$. It is a homeomorphism and seems to be differentiable. }
\label{fig1a}
\end{figure}

\begin{figure}[h]
  \psfig{file=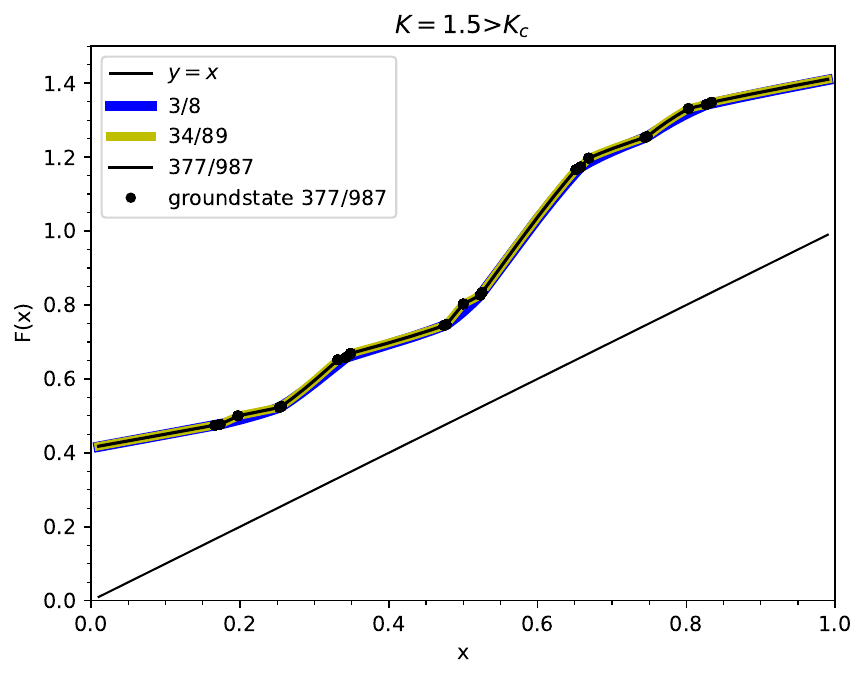,width=8.7cm,angle=-0}
  \caption{The same as Fig.~\ref{fig1a} for $K>K_c$ (pinned phase). $F$ remains a homeomorphism but the broken line aspect indicates that it is no longer differentiable. It is obtained by constrained minimization.}
\label{fig1b}
\end{figure}

Since it is  impossible to compute the ground state corresponding to an irrational rotation number $\rho$ numerically, one generally uses a sequence of rational approximants $r/s$ which converges to $\rho$. In this case, it is possible to obtain the ground state configuration $x_n$ as in section~\ref{constraints}. Then, one can draw $x_{n+1}$ as a function of $x_n$, i.e. $F$ for $s$ distinct points appropriately taken in $[0,1]$ (see the points in Fig.~\ref{fig1a} or Fig.~\ref{fig1b}).

In Fig.~\ref{fig1a} and \ref{fig1b}, we show the results for three rational numbers, $3/8$, $34/89$ and $377/987$ which, would the sequence be continued, converge to the irrational number $\rho=(3-\sqrt{5})/2$, for which the critical coupling of the Aubry transition is $K_c=0.9716...$ (see Ref.~\onlinecite{Greene} or \onlinecite{AubryQuemerais} and the discussion in section~\ref{Pascal}). In Fig.~\ref{fig1a}, we consider $K=0.5<K_c$ and the points defining $F$ become more and more dense as $s$ increases and define a continuous graph. On the other hand, for $K=1.5>K_c$ (Fig.~\ref{fig1b}), the points are no longer dense physically (even then for the last approximant used) and there is some arbitrariness in the definition of $F$ within the gaps.
Bangert uses a linear increasing interpolation between the points~\cite{bangert}: only the discrete points $x_n$ are physical and the points of the linear interpolation are not. As an alternative we can use a more physical construction with atoms at fixed ends.
We fix an atom at $x_0$ and look for a periodic state satisfying $x_{s}=x_0+r$, so that $x_{s}$ is also fixed. We let the other atoms adjust their positions $x_n$ such as to minimize the energy. In particular this gives $x_1$ which is plotted as a function of $x_0$, defining directly $F$. We observe in both figures that the subsequent $F$ of various rational approximants converge numerically onto a well-defined function $F$, which we define as the function $F$ corresponding to the irrational rotation number.
By construction, it satisfies $F(x+1)=F(x)+1$ and is found to be strictly increasing, as expected for a homeomorphism.

Note that if $x_0$ corresponds to a ground state position then the iteration of $F$ defines a ground state and $F$ is defined on $x_0$ and iterated points as in Bangert's work. Otherwise, the state obtained is not a ground state and has in general a higher energy. It corresponds to states inside the Peierls-Nabarro energy barrier, which is generally computed in this way~\cite{Peyrard}.

The difference in $F$ between $K<K_c$ and $K>K_c$ is clearly apparent.
For $K<K_c$ (Fig.~\ref{fig1a}), the points of the ground states are everywhere in $[0,1]$ and $F$ looks like a smooth function. For this small value of $K=0.5$, $F$ is actually quite close to $x+\rho$ (recall that $F(x)=x+\rho$ for $K=0$). For $K>K_c$ (Fig.~\ref{fig1b}), the points of the ground states are not everywhere and the rather straight portions of $F$ between them give a broken line aspect. This is even more apparent in the numerically-computed derivative of $F$ which appears to be discontinous (not shown). In fact, as will be discussed below, for $K>K_c$ Denjoy's theorem explains that $F$ cannot be smooth. 

We have illustrated the construction of Bangert type functions $F$ for the Frenkel-Kontorova model. They are homeomorphisms both for $K<K_c$ or $K>K_c$. However, they cannot be (and numerically are not indeed) C$^2$ diffeomorphisms for $K>K_c$.


\section{Coordinate changes and Denjoy constraints}

\begin{figure}[h]
  \psfig{file=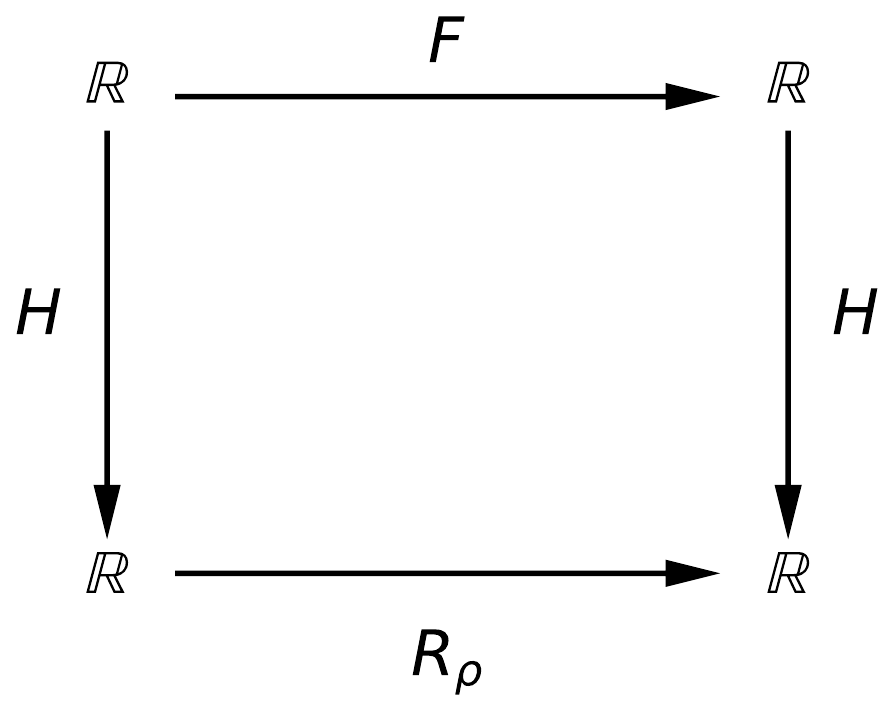,width=5.0cm,angle=-0}
  \caption{Change of variables $H$ and its representative diagram showing the equality (or semiconjugacy) between  $H \circ F$ (upper path) and $R_{\rho} \circ H$ (lower path) applied to any point of $\mathbb{R}$. $F$ is a lift of a circle homeomorphism and $R_{\rho}$ an irrational translation by $\rho$, $H$ is the semiconjugacy that may not be invertible.}
\label{figsemi}
\end{figure}

When $\rho$ is irrational and the orbit $\{ x_n \}$ obeys Eq.~(\ref{Rbis}), Poincaré has further shown that there exists a nonlinear coordinate change $H$ to a translation. It means that we can define new variables,
\begin{equation} \phi_n=H(x_n), \end{equation}
satisfying
\begin{equation}
\phi_{n+1} = \phi_{n}+\rho =  R_{\rho} (\phi_{n}),
\end{equation}
where $R_{\rho}(x)=x+\rho$ is a translation by $\rho$. The new variables $\phi_n$ therefore simply obey:
\begin{equation}
\phi_n=n\rho+\phi_0.
\end{equation}
The coordinate change $H$ is a \textit{continuous} and \textit{monotonically increasing} surjective function which satisfies $H(x+1)=H(x)+1$.
Mathematically $H$ is called a semiconjugacy and the change of coordinates is expressed by the equation 
\begin{equation} H \circ F= R_{\rho} \circ H, \label{semicon2}
\end{equation}
corresponding to the diagram of Fig.~\ref{figsemi}.
The homeomorphisms $F$ and $R_{\rho}$ are thus ``equivalent'' up to the coordinate change $H$. Transposed in angular variables $x_n~$mod~1 satisfying $x_{n+1}~\mbox{mod}~1=f(x_n~\mbox{mod}~1)$, we have also \begin{equation} h \circ f=R_{\rho} \circ h, \end{equation} where $h(x)=H(x)~$mod~1 is viewed as a function from the circle $S^1$ to itself. The relation of $h$ to $H$ is the same as that of $f$ to $F$, as explained at the end of section~\ref{effectiverotation}.

Poincaré showed that the coordinate change $H$ (or $h$) always exists for a homeomorphism $F$ with an irrational rotation number. However, in general, it is only a semiconjugacy  [Eq.~(\ref{semicon2})] but may not be a conjugacy. In a conjugacy $H$ is a homeomorphism. In a semiconjugacy $H$ is continuous and monotonically increasing but not necessarily strictly increasing (it may have some plateaux). In 1932, Denjoy emphasized the difference and showed that both are possible under some conditions on $F$.
Following Yoccoz~\cite{Yoccoz}, we explain below the difference betweeen the two possibilities and, furthermore, we emphasize the connections with the two phases occuring in the Aubry transition.

\subsection{$H$ injective - Ergodic regime - Sliding phase}

In the first case, suppose that $H$ is strictly increasing, hence injective. Then $H^{-1}$ exists:
\begin{equation}  F= H^{-1} \circ R_{\rho} \circ H,
\end{equation}
This is called a topological conjugacy between $F$ and $R_{\rho}$ by the homeomorphism $H$.
By iteration from an initial condition $x_0$,
\begin{eqnarray}
x_n &=& F^{(n)}(x_0) \\ &=& H^{-1} \circ R_{\rho}^n \circ H (x_0),
  \end{eqnarray}
we therefore have the following solution:
\begin{eqnarray}
x_n  &=& H^{-1}(n\rho+\phi_0), \label{sol}
  \end{eqnarray}
where $\phi_0=H(x_0)$ is a phase.  The function $H^{-1}$ is thus Aubry's hull function~\cite{Aubry} and is continuous in the present case.  Since $\rho$ is irrational, $n\rho +\phi_0~$mod~1 is dense in $[0,1]$ and $x_n$~mod~1 is also dense because $H$ is a homeomorphism. There is a unique invariant set $[0,1]$ for the angular variables and the phase is ergodic. The minimal energy states are that of the sliding phase: one can translate them continuously by changing $\phi_0$ (or $x_0$), giving an emergent continuous symmetry which is not present in the model.

\begin{figure}[h]
      \psfig{file=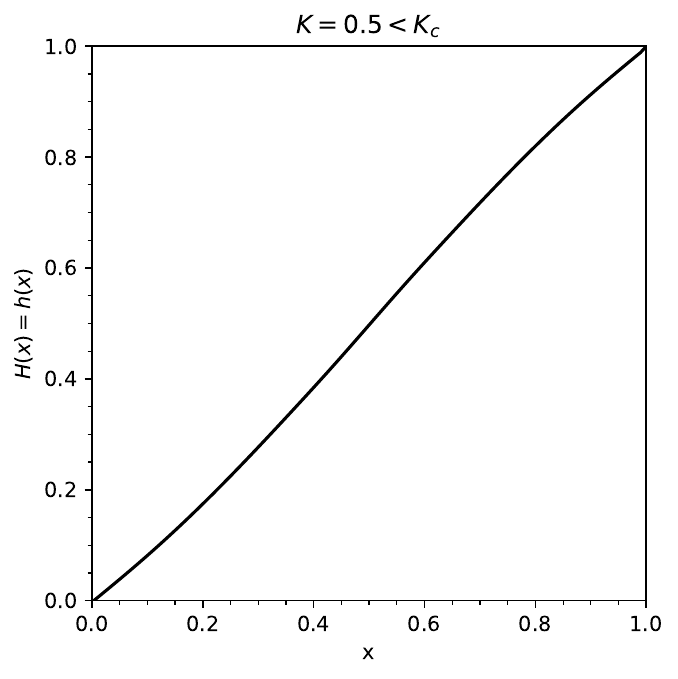,width=8cm,angle=-0}
        \psfig{file=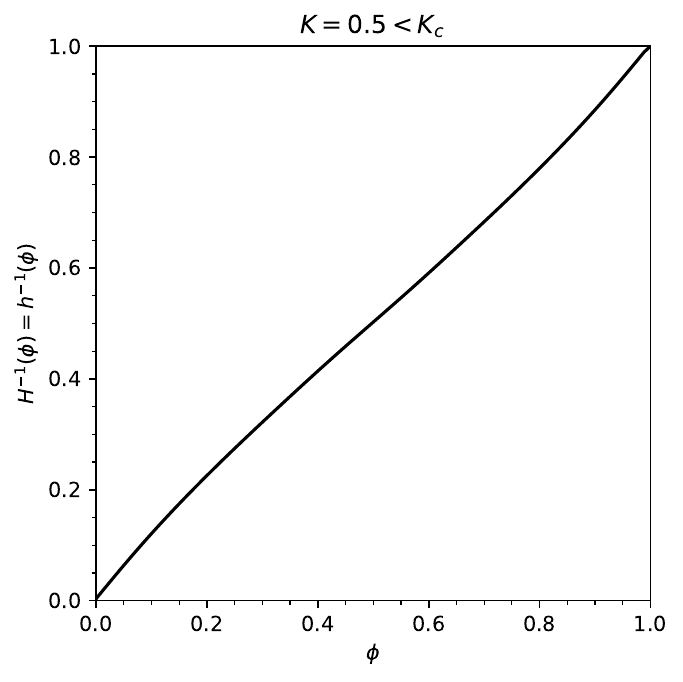,width=8cm,angle=-0}
  \caption{(Top) Topological conjugacy function $H$ for an irrational $\rho$ and $K=0.5<K_c$. $H$ is a homeomorphism. (Bottom) Inverse function $H^{-1}$ corresponding to Aubry's hull function [Eq.~(\ref{sol})] which is continuous for $K<K_c$.}
\label{fig3}
\end{figure}

In Fig.~\ref{fig3}, $H$ and $H^{-1}$ are computed numerically for $K=0.5<K_c$. This is done  by using Eq.~(\ref{sol}) (and its inverse) and a numerically-obtained ground state of the Frenkel-Kontorova model, $x_n$ (with $\rho=377/987$).
The function $H$ thus obtained is the conjugacy corresponding to the function $F$ shown in Fig.~\ref{fig1a}. We see indeed that $H$ is a homeomorphism. It is slightly deformed from the identity function $H(x)=x$ at $K=0$, but remains continuous and \textit{strictly} increasing. The hull function $H^{-1}$ is also continuous.

\subsection{$H$ non-injective - Cantor function - Pinned phase}

In the second case, $H$ is not injective and $H^{-1}$ does not exist, at least not as a function in the usual sense. Therefore, $H$ is no longer a homeomorphism so that there is no topological equivalence between the set of possible values $x_n$~mod~1 for the minimal energy states and the circle $S^1$. Such a case turns out to exist, as emphasized by Denjoy and illustrated below by Fig.~\ref{fig4}.

Since $H$ is monotonically increasing, but not strictly in this case, $H^{-1}$ can be seen as a multivalued function. For any $\phi$, $H^{-1}(\phi)$ may be either an interval or a point. In both cases, one can define functions $x_+$ and $x_-$ such that $H^{-1}(\phi)=[x_-(\phi),x_+(\phi)]$, with $x_-(\phi)=x_+(\phi)$ if the interval is reduced to a single point. By definition, for any point $x \in [x_-(\phi),x_+(\phi)]$, we have $H(x)=\phi$, i.e. $H$ is constant on any of these intervals and all these points $x$ may be seen as ``inverses'' of $\phi$ by the multivalued function $H^{-1}$.

\begin{figure}[h]
\psfig{file=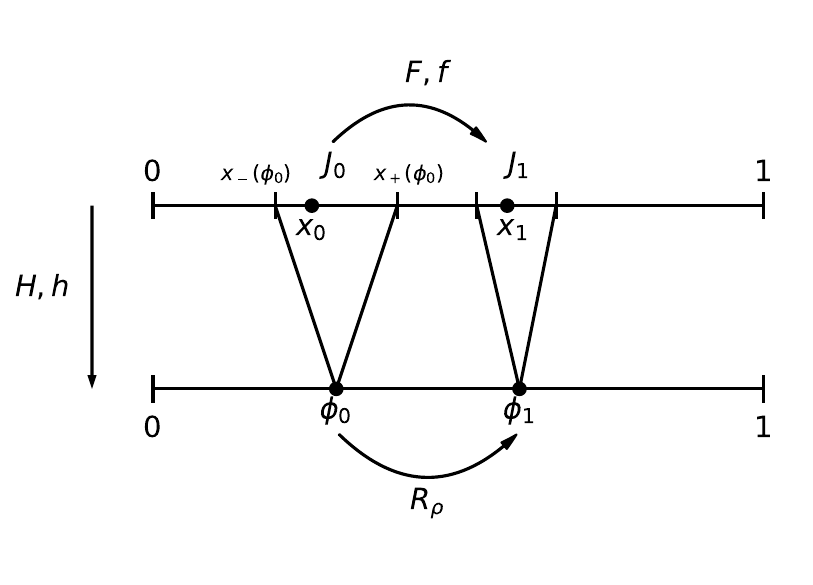,width=8.5cm,angle=-0}
\caption{Illustration of a non-injective semiconjugacy $H$ (see Fig.~\ref{fig4} for a graph of $H$). $H^{-1}(\phi_0)$ is a finite interval.}
\label{fig2}
\end{figure}

\begin{figure}[h]
      \psfig{file=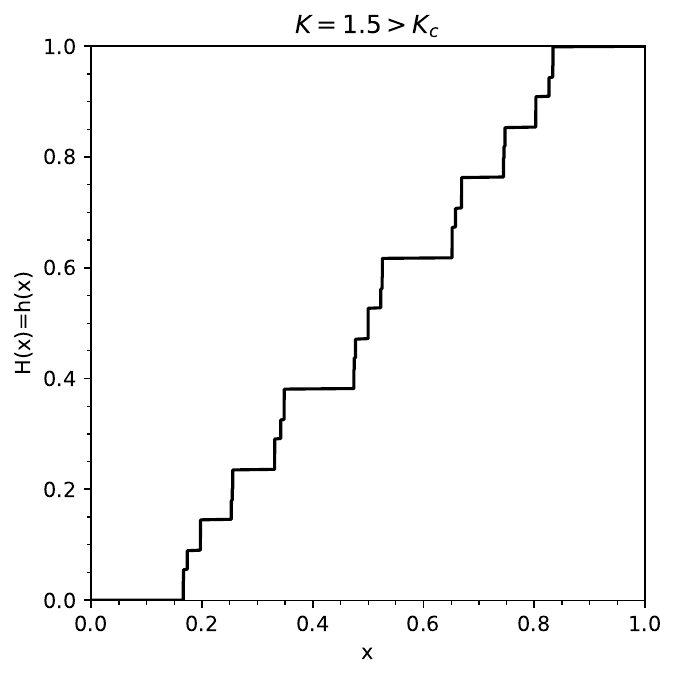,width=8cm,angle=-0}
        \psfig{file=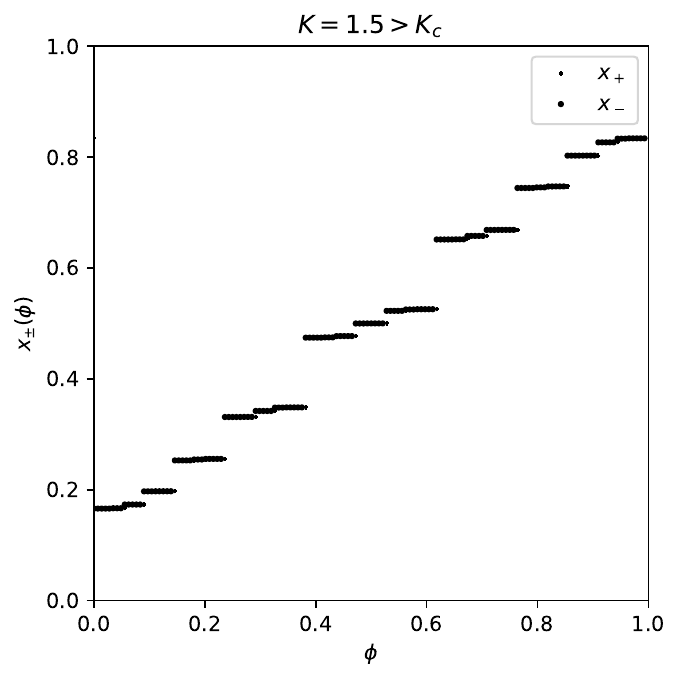,width=8cm,angle=-0}
  \caption{(Top) Semiconjugacy function $H$ for an irrational $\rho$ and $K=1.5>K_c$. $H$ is not a homeomorphism: it is a continuous Cantor function. The seemingly vertical parts are in fact made of many small plateaux. (Bottom) ``Inverse'' multivalued function $H^{-1}$. Here two special ``inverses'' $x_{\pm}$ [Eqs.~(\ref{xmenv})-(\ref{xpenv})] are given (recall that $H(x_{\pm}(\phi))=\phi$). They correspond to Aubry's hull functions and are discontinuous in this case and give two recurrent ground states. $x_+$ and $x_-$ are almost undistinguishable in the figure. Another ``inverse'' is given in Fig.~\ref{fig8}.  }
\label{fig4}
\end{figure}

In the present case, there exists \textit{at least} one point $\phi_0$ for which $H^{-1}(\phi_0)$ is an interval. We denote the corresponding non-empty open interval by $J_0=]x_-(\phi_0),x_+(\phi_0)[$ (see Fig.~\ref{fig2}). In fact there exists many of them. We suppose with no loss of generality that $J_0$ belongs to $[0,1]$ and discuss the dynamics modulo one (i.e. in $[0,1]$), i.e. by using $f$ and $h$ instead of $F$ and $H$.  The image of $J_0$ by the circle homeomorphism $f$ is also an interval $J_1$; for any point $x_0$ in $J_0$, $x_1=f(x_0)$ is in $J_1$. By definition of the semiconjugacy $h$, we have $h(x_1)=h \circ f(x_0)=R_{\rho} \circ h(x_0)=R_{\rho} \phi_0=\phi_0+\rho=\phi_1 \neq \phi_0$, therefore $h(J_1)=\phi_1$ the image of $\phi_0$ by the rotation $R_{\rho}$ (see Fig.~\ref{fig2}). $J_1$ is a new such interval. Similarly, $J_n \equiv f^{(n)}(J_0)$ with $h(x_n)=\phi_n$ for all points $x_n$ in $J_n$ are also new such intervals.
    
For that reason, any non empty interval $]x_-(\phi),x_+(\phi)[$ on which $H$ is constant and equal to $\phi$ defines a series of other intervals by the dynamics and is thus called a ``wandering'' interval. We denote by $W$ the union of all wandering intervals, which contains at least $J_0$ and all $J_n$, as we have seen and, possibly, others.
    The following properties are discussed for example by Yoccoz~\cite{Yoccoz}.
    
\begin{itemize}
\item The wandering intervals do not overlap because $h$ is constant on each interval: if there was a common point in two wandering intervals they would form a single wandering interval. Thus any two wandering intervals either coincide or are disjoint.
\item They are dense: if they were not, there would be a finite interval $I$ which does not contain a wandering interval. The image of $I$ by $h$ would be a finite interval $h(I)$, it cannot be reduced to a point because $I$ itself would then be a wandering interval.
  But in $h(I)$ there must be some points of the form $R_{\rho}^k \phi_0=\phi_k$ (we have seen that $\phi_0$ necessarily exists), because those $\phi_k$ are dense in $[0,1]$ for irrational $\rho$. That contradicts the absence of wandering intervals in $I$.
\end{itemize}

The union of wandering intervals $W$ is invariant by $f$: a trajectory which starts at a point $x_0$ in $W$ in a given wandering interval $J_0$ remains in $W$. It is not recurrent because its images belong to successive $J_n$
 intervals which are all disjoint. The complement $C= S^1 \backslash W$ is also invariant by $f$: it consists of trajectories which start in $C$ and remain in $C$. $C$ is in fact a Cantor space (a compact totally discontinuous space with no isolated points):
    \begin{itemize}
      \item Since $W$ is open, its complement $C$ is a closed subset of $S^1$ and any closed subset of a compact space is compact.
      \item $C$ is totally discontinuous: it contains no finite interval because the union of the wandering intervals $W$ is dense. \item $C$ has no isolated points: if there was an isolated point $x$, there would be a wandering interval on each side of $x$. The function $h$ is constant on each wandering interval and would have two distinct values on those two wandering intervals.  But this is not possible because $h$ is continuous.  Thus $h(x)$ would be equal to that common value and hence the union of the two intervals plus the point $x$ would form a unique wandering interval which is a contradiction so that $x$ cannot be isolated.
\end{itemize}

If we return to the dynamics in $\mathbb{R}$ and consider $H$, we have the following properties:
\begin{itemize}
  \item $H$ is constant on each wandering interval, the collection of which are dense in $[0,1]$ and then in $\mathbb{R}$ by using $H(x+1)=H(x)+1$. So it is constant almost everywhere.
  \item $H$ is also continuous and increasing.
\end{itemize}
So $H$ is a Cantor function (a complete Devil's staircase). Note that it is almost everywhere differentiable with zero derivative. Thus the function $H$ is equal to its singular part (in the language of Rudin~\cite{Rudin}). In particular, $H$ is not equal to the integral of its derivative. 

Iterating $F$ in the present case gives two types of trajectories depending on the initial condition $x_0$. They are either in $W$ and those are not recurrent or in $C$ and those are recurrent.
At this point of the discussion, one cannot tell to which set the minimal energy state belongs. The initial condition $x_0$ (with corresponding phase $\phi_0=H(x_0)$) is chosen in a given interval $[x_-(\phi_0),x_+(\phi_0)]$, with three distinct possibilities: either $x_0=x_{+}(\phi_0)$ or $x_0=x_{-}(\phi_0)$ at one or the other end of the interval (those two points may be the same if the interval is reduced to a point) or $x_0$ strictly inside the interval (if the interval is not reduced to a point). The possible orbits (and minimal energy states) $x_n$ then follow:
\begin{eqnarray}
  x_n &=& x_-(n\rho+\phi_0), \label{x-} \label{xmenv} \\ 
  x_n &=& x_+(n\rho+\phi_0), \label{x+} \label{xpenv} \\
  x_n &=& x(n\rho+\phi_0),  \label{xenv}
\end{eqnarray}
where $x$ is also an ``inverse'' of $H$ in the sense that $H(x(\phi))=\phi$.
The first two orbits belong to the Cantor set $C$ and are recurrent.  The third orbit belongs to $W$. Since the dynamics passes only once in each $J_n$, those states are not recurrent. Note that in general there may be several orbits starting from different wandering intervals with $\phi$ independent of $\phi_0$ and not related by the dynamics~\cite{AubryGosso}.

We claim, with the construction of $F$ of section~\ref{num}, that the three possibilities [Eqs.~(\ref{xmenv}), (\ref{xpenv}) and (\ref{xenv})] correspond to different physical states.  The first two correspond to two ground states and $x_{\pm}$ are Aubry's \textit{discontinuous} hull functions. The ground states are pinned since the discontinuities reflect the fact that $x_n$ never belong to any open interval $J_n$ which are forbidden regions. The third one corresponds to a state with a higher energy inside the Peierls-Nabarro barrier (see section~\ref{num}). By varying continuously $x_0$ from that of the first ground state to that of the second, the orbits are deformed and the energy increases and follows the energy barrier. 
Thus, the two distinct invariant dynamical sets $C$ and $W$ describe two types of physical states with different energies.

\begin{figure}[h]
  \psfig{file=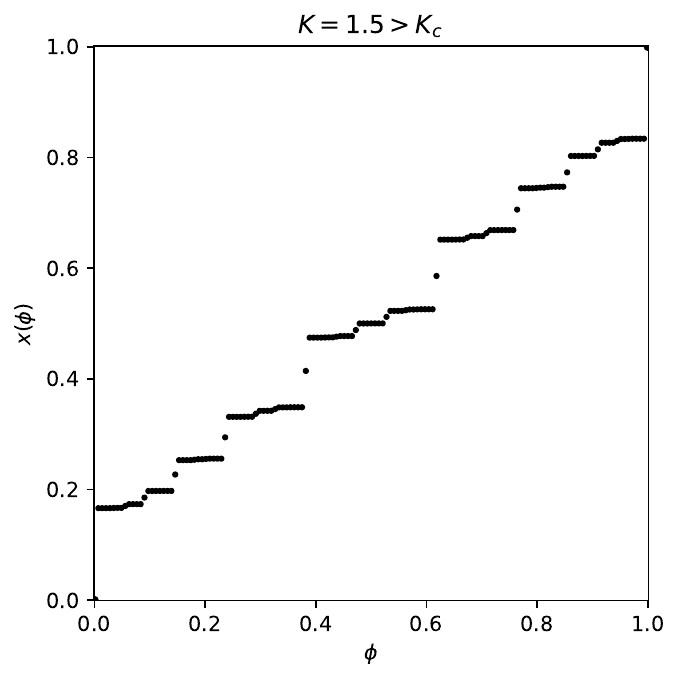,width=8cm,angle=-0}
  \caption{The function $x$ [Eq.~(\ref{xenv})] describes a higher energy nonrecurrent state in the Peierls-Nabarro energy barrier, intermediate between $x_+$ and $x_-$. It is also an ``inverse'' of $H$ (given in Fig.~\ref{fig4}, top) in the sense that $H(x(\phi))=\phi$. Note that one atom, say the atom 0, is fixed at $x_0=x(0)=0$, i.e. at the maximum of the potential.}
\label{fig8}
\end{figure}

In Fig.~\ref{fig4} (top) we give the function $H$ computed numerically as above, but for $K>K_c$.
It is the semiconjugacy corresponding to the function $F$ given in Fig.~\ref{fig1b} and is expected to be a Cantor function. The function $H$ obtained numerically is not strictly speaking a Cantor function because its self-similarity stops at some scale which depends on the rational approximant (here $\rho=377/987$).
The hull functions $x_{\pm}$ are given in Fig.~\ref{fig4} (bottom). They are discontinuous,
corresponding to the breaking of analyticity. In the figure we plotted both $x_+ $ and $x_-$ but if we had plotted only one of them, the figure would have looked almost the same, although $x_+$ and $x_-$ differ at each discontinuity.
 The function $x$ [Eq.~(\ref{xenv})] is given in Fig.~\ref{fig8} and is intermediate between the previous two. It is a higher energy state which is not recurrent because it consists of isolated points within the gaps (here $x_0=0$ was chosen to be at the middle of the largest gap which corresponds to a maximum of the potential).


\subsection{The Aubry transition in the Frenkel-Kontorova model}
\label{Pascal}

The transition from conjugacy (Fig.~\ref{fig3}, top) to semiconjugacy (Fig.~\ref{fig4}, top) corresponds to Aubry's breaking of analyticity (from Fig.~\ref{fig3}, bottom to Fig.~\ref{fig4}, bottom). In the Frenkel-Kontorova model, it depends on the parameter $K$, the amplitude of the periodic potential. Intuitively, one can easily understand that when $K$ is large, no atom in a ground state can remain near the maxima of the potential. According to the previous discussion, this implies that for an incommensurate ground state (irrational $\rho$), the set of $x_n$~mod~1 cannot be dense on the circle. We are then in the case of semiconjugacy, and the $x_n$~mod~1 occupy positions on a Cantor set. Note that it is not at all physically obvious that, not only can atoms not be present near the maxima of the potential, but they also cannot be present in any of the infinitely many wandering intervals, which are densely spaced over the period of the potential. This is the pinned phase.

Conversely, when $K$ is sufficiently small, the elastic part of the Frenkel-Kontorova model dominates the energy balance, and the presence of atoms close to the maxima of the potential is allowed. We are in the case of conjugacy, and the set of positions $x_n$~mod~1 is dense on the circle. This is the sliding phase.

A simple and more precise argument can be given~\cite{Aubry7D}. The inequality Eq.~(\ref{inequality}) shows that the atomic positions are constrained by the homeomorphism $F$. If we apply it for $m = n-1$, we get $\vert x_n - x_{n-1} - \rho \vert < 1$, hence $\vert x_{n+1}+x_{n-1}-2x_n \vert<2.$ By taking into account the extremalisation equation [Eq.~(\ref{firstderivative})], we obtain,
$$\vert x_{n+1}+x_{n-1}-2x_n \vert= \frac{K}{2 \pi} \vert \sin{(2\pi x_n)} \vert <2.$$
If $x_n$~mod~1 can take all values in $[0,1]$, as in the sliding phase, the inequality is impossible to satisfy whenever $K > 4\pi$, i.e. the equilibrium of forces cannot be satisfied. The sliding phase cannot exist. All incommensurate phases are therefore pinned for those $K$, and the atoms, regardless of the irrational rotation number $\rho$, occupy positions on a Cantor set. This shows that for every $\rho$, $K_c=K_c(\rho) \leq 4\pi$.
Refinements of this bound were obtained by Aubry~\cite{Aubry7D} and also by MacKay and Percival~\cite{MacKayPercival}, who showed that $K_c \leq \frac{63}{64} \approx 0.9843 $. Numerically, Greene~\cite{Greene} found an upper bound very close to $0.9716$. This numerical value was obtained for the golden mean $\rho=(1+\sqrt 5)/2$ (and is the same for related numbers such as the one used above $(3-\sqrt{5})/2$). Greene argued that because it is the irrational number least easily approximated by rationals, this corresponds precisely to the most favorable conditions for the KAM theorem to be applied: the problem of small denominators is, in a way, minimal for the golden ratio (and related numbers). Then, for any other irrational number $\rho$, there exists a critical value $K_c < 0.9716\dots$ above which the structure is pinned and below which the KAM theorem applies and the structure is sliding. At this stage of our current knowledge, it is very difficult to prove the convergence of the series involved in the KAM theorem for a given irrational number. However, good approximate values of $K_c$ for different $\rho$ have been obtained numerically\cite{Schmidt,MacKayStark2}.

\section{Conclusion}

The Aubry transition can be viewed as a topological phase transition corresponding to a change in the topology of the ground state manifold. In the sliding phase the ground state variables $x_n$~mod~1 are dense in $[0,1]$ and the manifold is a whole circle. The ground states have therefore an emergent continuous symmetry that the model does not have. In the pinned phase, the same variables belong to a Cantor space with infinitely many open gaps and forbidden regions. This phase does not have any longer the continuous symmetry. This transition from a circle to a Cantor set (or cantorus) is encoded in the coordinate change $H$ that describes the ground states in terms of rotation, which is either a topological conjugacy (homeomorphism) or a semiconjugacy.  This is because the ground states are constrained by circle homeomorphisms $F$, which is a consequence of the properties of the model.  Denjoy recognized that there are two possibilities depending on the regularity of $F$. For the Aubry transition of the Frenkel-Kontorova model for example, the regularity properties of $F$ are not known but can thus be inferred from Denjoy's theorem. The ones that we have numerically constructed perfectly agree with that viewpoint. For $K<K_c$, $F$ appears to be smooth (Fig.~\ref{fig1a}) and $H$ is a homeomorphism (Fig.~\ref{fig3}). For $K>K_c$, $F$ appears to be non-differentiable (Fig.~\ref{fig1b}) and $H$ is a Cantor function (Fig.~\ref{fig4}).
    This schematic and qualitative discussion of the two cases on the basis of the existence of the homeomorphism $F$, as emphasized by Bangert, is then interesting in the physical context and may open new ways to prove the existence of the Aubry transition in other models than the Frenkel-Kontorova model: while as we have recalled the form of the energy (\ref{newenergy}) necessarily implies the existence of $F$, other models should also work but the exact ingredients are not known.
    However this anachronistic point of view hides some important results: the proof of existence of the sliding phase (resp. pinned phase) in the Frenkel-Kontorova model for small $K$ (resp. large $K$) and the precise quantitative determination of $K_c$ as a function of $\rho$. In particular the existence of the sliding phase protected by the KAM theorem for $K \neq 0$ is not discussed.

However, it emphasizes more generally
the issue of the relationship between the smoothness of the homeomorphism $F$ and that of $H$, first illustrated by Denjoy. It has been further studied mathematically since then. For example the KAM theorem tells us that if $F$ is analytic (for example) and close enough to an irrational rotation, then $H$ is analytic. Arnold's conjecture (proved by Herman~\cite{Herman} and Yoccoz~\cite{Yoccozth}) is that for all analytic $F$ (not necessarily close to a rotation) and Diophantian rotation numbers (Yoccoz proved the theorem for a less restrictive class of irrational numbers~\cite{Yoccozth1}), $F$ is conjugate to a rotation and $H$ is analytic. While the Aubry transition is related to a change of regularity of $F$ from (one would claim) smooth to non-differentiable (with consequence that $H$ changes from a homeomorphism to a non-bijective function), thus, more generally, one may wonder what are the conditions in the energy model to have other types of regularity changes, leading possibly to different forms of Aubry-like transitions.

\begin{acknowledgments}
  This paper was written in honor of Serge Aubry's 80th birthday. One of us (P.~Q.) wishes to express his gratitude to him for having opened the doors to theoretical physics. Thirty-five years later, the period of collaboration with Serge remains one of his best scientific and personal memories.

  We would like also to thank a referee for pointing out a generalization of the concept of conjugacies and circle maps in terms of the ``parametrization method''~\cite{parametrization} which may be helpful in higher dimensional cases.
\end{acknowledgments}


\begin{thebibliography}{99}
\bibitem{acc} Energy terms that violate a given symmetry may be irrelevant in renormalization group iterations.
\bibitem{ft1} $E$ can be seen as the action of a dynamical system in discrete time (e.g. the standard map) and $L$ its Lagrangian.
\bibitem{notePropertiesL} The function $L(x,y)$ is generally assumed (see Aubry-Le Daeron~\cite{aubry_ledaeron})  to have a lower bound, a periodicity $L(x+1,y+1)=L(x,y)$ and to be a two times differentiable function satisfying $\frac{\partial^2 L}{\partial x \partial y}<C<0$ (where $C$ is a negative constant). For more general conditions and the connection with the dynamical ``twist'' condition, see Bangert~\cite{bangert}.
\bibitem{bijection0} By definition, any point of a minimal energy state $x_n$ has at least one antecedent $x_{n-1}$, so that $F$ is surjective. But it cannot have more than one antecedent, this would mean that there are two minimal energy states crossing at that point $x_n$, which is impossible for the ensemble of states considered. $F$ is therefore also injective. $F$ is strictly increasing because whenever, for two trajectories of the ensemble of translated and renumbered states, $x_n<x_n'$ implies $x_{n+1}<x_{n+1}'$, because the ensemble is totally ordered.
\bibitem{bijection} The precise meaning of angular variables being arranged ``in the same way'' is that the actual real-valued variables satisfy the following property. There is an increasing bijection between the ensemble of points $q \rho+p$ where $p$ and $q$ are integers and the ensemble of points $x_q+p$. It means that whenever $q\rho+p<q'\rho+p'$ then $x_q+p<x_{q'}+p'$. For a proof, see for example~Ref.~\onlinecite{Yoccoz} or proposition 11.2.4 page 395 of Ref.~\onlinecite{Katok}.
\bibitem{AubryGosso} Such additional orbits are important in other models than the original Frenkel-Kontorova model. They appear in Ref.~\onlinecite{Aubrydisco} as different families of discommensurations, but will not be discussed further here.
\end{thebibliography}
\end{document}